\documentclass[a4paper]{article}

\usepackage[english]{babel}
\usepackage[utf8x]{inputenc}
\usepackage[T1]{fontenc}

\usepackage[a4paper,top=3cm,bottom=2cm,left=3cm,right=3cm,marginparwidth=1.75cm]{geometry}

\usepackage{amsmath}
\usepackage{graphicx}
\usepackage{subcaption}
\usepackage{color, colortbl}
\usepackage[colorinlistoftodos]{todonotes}
\usepackage[colorlinks=true, allcolors=blue]{hyperref}

\usepackage{booktabs}

\title{The Skipping Behavior of Users of Music Streaming Services \\
  and its Relation to Musical Structure}

\author{Nicola Montecchio, Pierre Roy, François Pachet \\ Spotify}

\begin{document}
\date{}
\maketitle

\begin{abstract}

The behavior of users of music streaming services is investigated from
the point of view of the temporal dimension of individual songs;
specifically, the main object of the analysis is the point in time
within a song at which users stop listening and start streaming
another song (``skip'').  The main contribution of this study is the
ascertainment of a correlation between the distribution in time of
skipping events and the musical structure of songs.  It is also shown
that such distribution is not only specific to the individual songs,
but also independent of the cohort of users and, under stationary
conditions, date of observation.  Finally, user behavioral data
is used to train a predictor of the musical structure of a song
solely from its acoustic content; it is shown that the use of
such data, available in large quantities to music streaming services,
yields significant improvements in accuracy over the customary fashion
of training this class of algorithms, in which
only smaller amounts of \mbox{hand-labeled} data are available.

\end{abstract}

\section{Introduction}

Since the advent of online music streaming services, people have been
able to easily access millions of songs on demand. As a consequence of
this abundance, novel listening behaviors have emerged: departing from
the passive, concentrated listening practice that is typical of media
such as LPs, today people tend to listen to music in a much more
frantic way than before.

The central object of interest of this paper is the behavior of users
regarding ``skipping'': the act of interrupting a song in order to
listen to the next song in the music queue (the queue possibly being the
song's album, a playlist in which the song figured, or a sequence of
songs proposed by the recommendation engine of the streaming service).

It is our opinion that skipping is a crucial feature in understanding
modern listening behaviors. For the first time in the history of
musicology, researchers can systematically collect and analyze massive
amounts of data about music listening behavior.  Streaming services
are only one of many possible contexts in which music is consumed; as
such, one must be aware that any observed behaviour is not necessarily
generalizable to other situations (e.g., it is safe to assume that
skipping behaviour in the context of listening to LPs is radically
different from the online music streaming case described in this
paper); nonetheless, online streaming services represent nowadays the
preferred music consumption mechanism.
Statistical analysis of user skipping behavior in time yields indeed
fascinating information about how people listen and react to music.

We start by investigating the specificity of skipping behavior with
respect to songs, in particular with respect to its consistency in the
case of data collection on different dates and in different regions.
We then identify a connection, which to the best of our knowledge has
not been observed before, between skip behaviour and musical structure
(the segmentation of a musical work into musically relevant sections,
such as ``intro'', ``verse'', ``chorus''): listeners are more likely
to skip a song right after a change between musical sections.
Subsequently, we turn our attention to the task of predicting musical
structure from acoustic content (i.e., from the raw audio waveform of
a recording), a well known problem in the research community, commonly
referred to as \emph{structural segmentation}: we propose a novel
approach that makes use of skipping behavior in order to train a
prediction algorithm in a \mbox{semi-supervised} way, by exploiting
data that can be extracted automatically in large quantities and is
directly related to the perception of music by users. We finish by
discussing how future lines of inquiries that stem from this study,
aimed at a better understanding of user reaction, could potentially
lead to the development of compositional tools aimed at improving the
reception of music by audiences.

\section{General Statistical Aspects of User Behavior}
\label{sec:generaltrends}

\begin{figure}

\centering

  \begin{subfigure}{0.49\textwidth}
    \centering
    \includegraphics[width=1.0\textwidth,trim={0.1cm 1.2cm 0 0},clip]{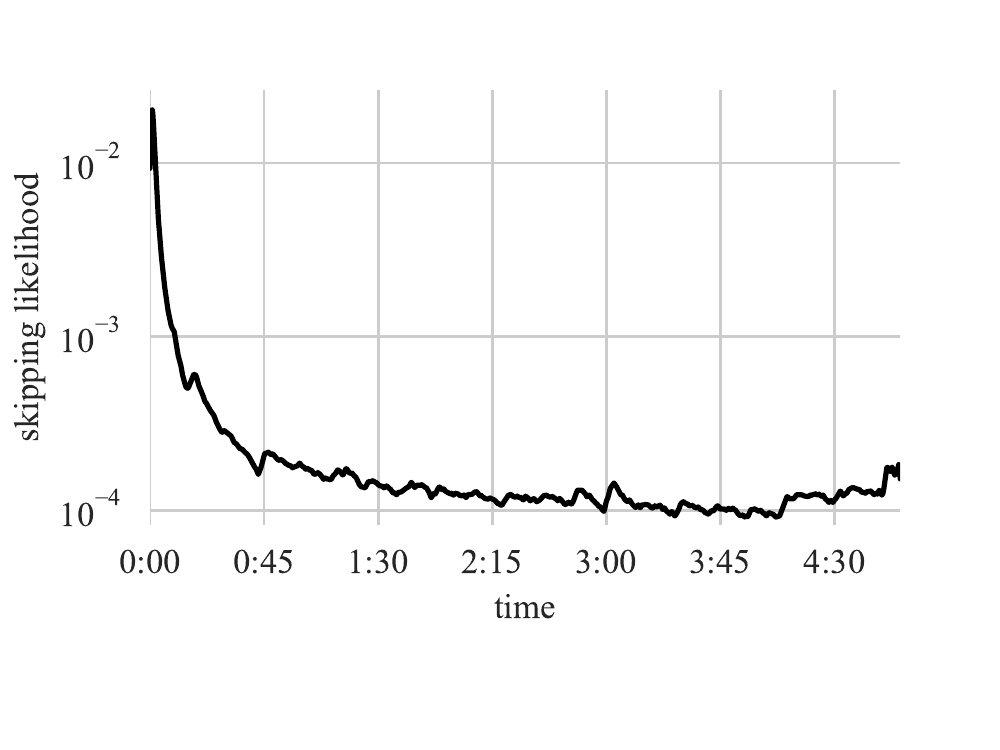}
    \caption{ \label{fig:spbadbunny}   An individual song. }
  \end{subfigure}%
  ~
  \begin{subfigure}{0.49\textwidth}
    \centering
    \includegraphics[width=1\textwidth,trim={0.9cm 0.1cm 1.4cm 0},clip]{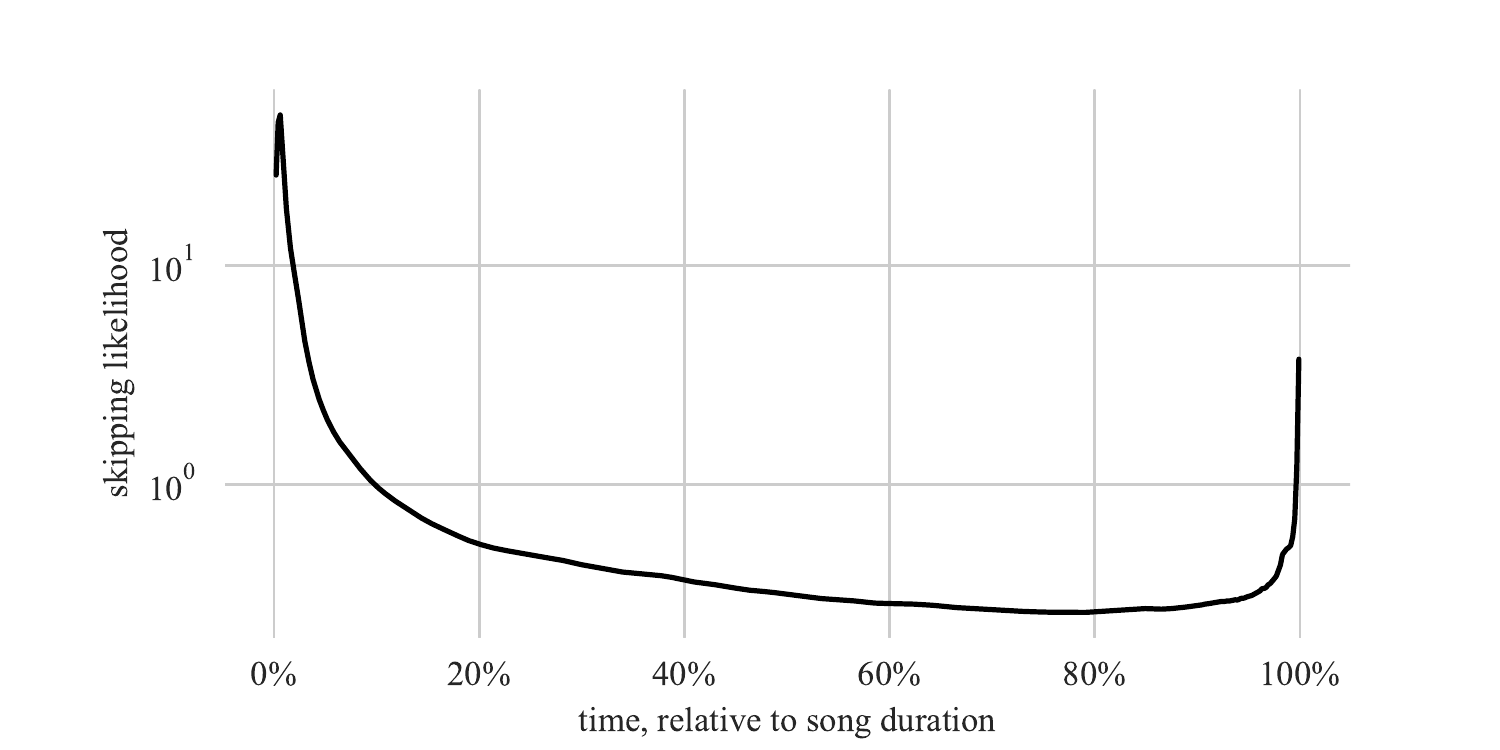}
    \caption{\label{fig:percentageprofile} Aggregated streams of all songs. }
  \end{subfigure}

  \caption{ \label{fig:skipprofileexample} Skip profiles of individual songs and aggregate behavior.  }

\end{figure}

The principal object of study of this paper is the distribution (histogram) of the
points in time at which users stop listening to a track,
which will be referred to as the track's ``\emph{Skip Profile}''; a typical example
 is depicted in Figure~\ref{fig:spbadbunny}.

Visual inspection of Skip Profiles suggests a possible interpretation
as the superposition of a general trend (pictured in
Figure~\ref{fig:percentageprofile}, obtained through the aggregation
of streaming data over the entire catalog) and a residual signal in
which of peaks concentrate at specific points in time.
In this Section, we investigate the
collective behavior of users on the platform, as well as the specificity
of skip profiles to songs and their consistency in time and across
geographical regions.

\subsection{Previous Work}

Previous research has addressed the issue of modeling music listening;
in particular, many works investigated why certain songs become more
popular than others. \cite{Salganik854}~shows that the non-uniformity
of subjective preference maybe largely explained by so-called
\emph{cumulative advantage}, therefore rendering a priori prediction
of popularity rather pointless: music hits are inherently
unpredictable, due to social pressure. Nonetheless, many works
addressing ``Hit Song Science'' have been published
(e.g. \cite{HerremansHSS}), attempting to predict the popularity of a
new song based on features automatically extracted from its acoustic content;
criticism of such line of works include \cite{PachetRoyHSS}. All these
studies, however, consider songs in their entirety, and do not
consider the impact of listeners from a temporal dimension.

A different line of research literature is concerned with the temporal
aspect of user responses to musical stimuli. The object of the
multiple experiments presented in~\cite{belfi2018rapid} is to
``identify the amount of time necessary to make accurate aesthetic
judgments''; in this study the authors argue that such time is around
750ms.  Most skipping activity in the context of a commercial music
streaming service also occurs in the very first few seconds of
listening, thereby suggesting a similar ability of users to very
quickly express (negative) aesthetic judgments: preliminary analysis
of Skip Profiles~\cite{LamereSkips}, averaged on millions of listeners
and billions of plays obtained from the commercial streaming service
Spotify, identifies a ``steep drop off in listeners in the early part
of a song, when most listeners are deciding whether or not to skip the
song''. It must be pointed out that the context of~\cite{LamereSkips}
(and of this paper) -- behavior of generic users in unspecified
listening contexts -- is radically different from the carefully
designed and controlled experimental conditions
of~\cite{belfi2018rapid}; nonetheless, the findings of both are in
agreement.

Related work done on large scale music listening behavior data
includes the analysis of \emph{scrubbing} behavior~\cite{LamereDrops},
that is the practice of moving the cursor \emph{within} the song in order to
search for, and listen to specific parts. The author showed how such data can be used
to identify segments of particular interest in songs: instrumental
solos, particularly dramatic moments, and, within the genre of
electronic dance music, the ``drop''.

\subsection{Average Behavior}

\begin{figure}

  \begin{subfigure}{0.49\textwidth}
    \centering
    \includegraphics[width=1\textwidth,trim={1cm 0.1cm 1.7cm 0},clip]{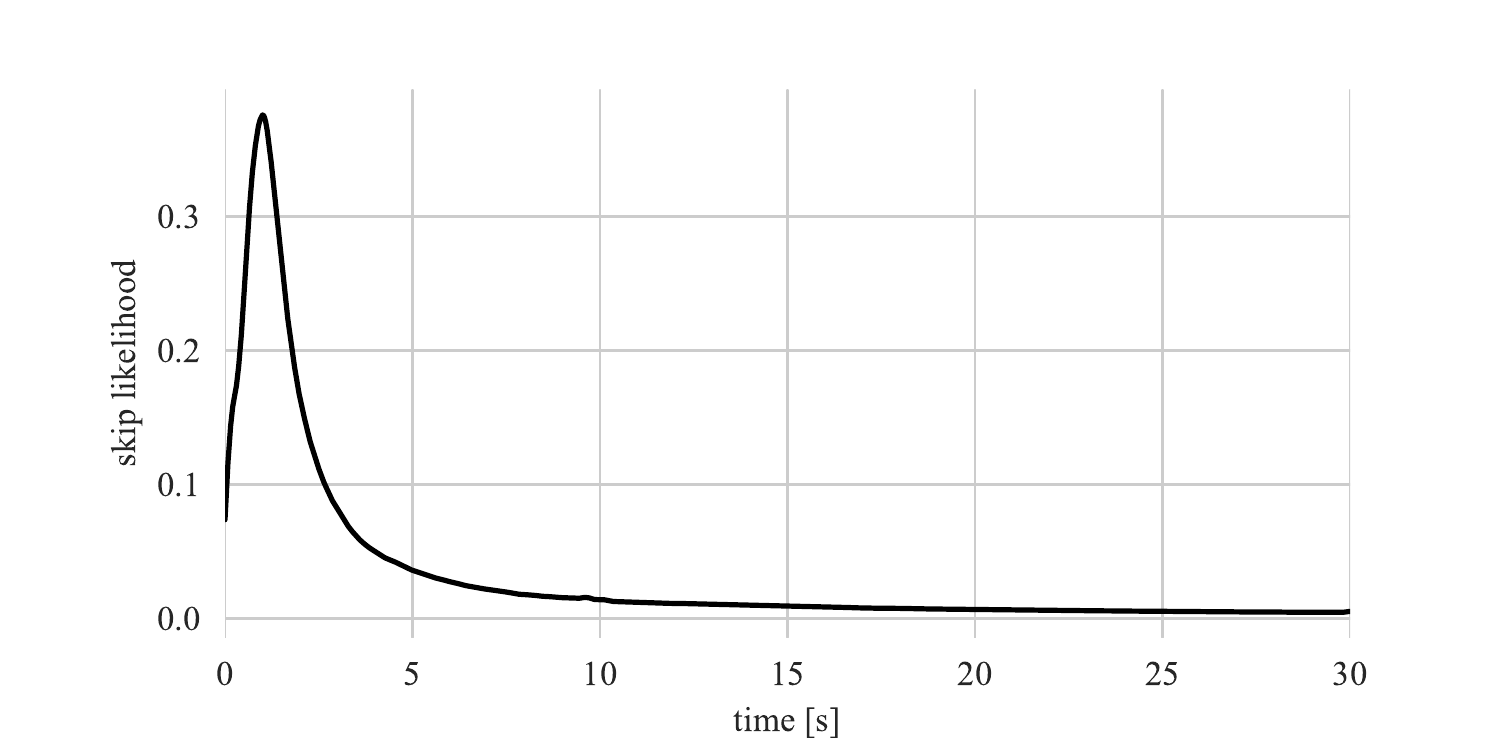}
    \caption{\label{fig:skipprofile30s} Skipping likelihood at the beginning of a song.}
  \end{subfigure}
  ~
  \begin{subfigure}{0.49\textwidth}
    \centering
    \includegraphics[width=1\textwidth,trim={1cm 0.1cm 1.7cm 0},clip]{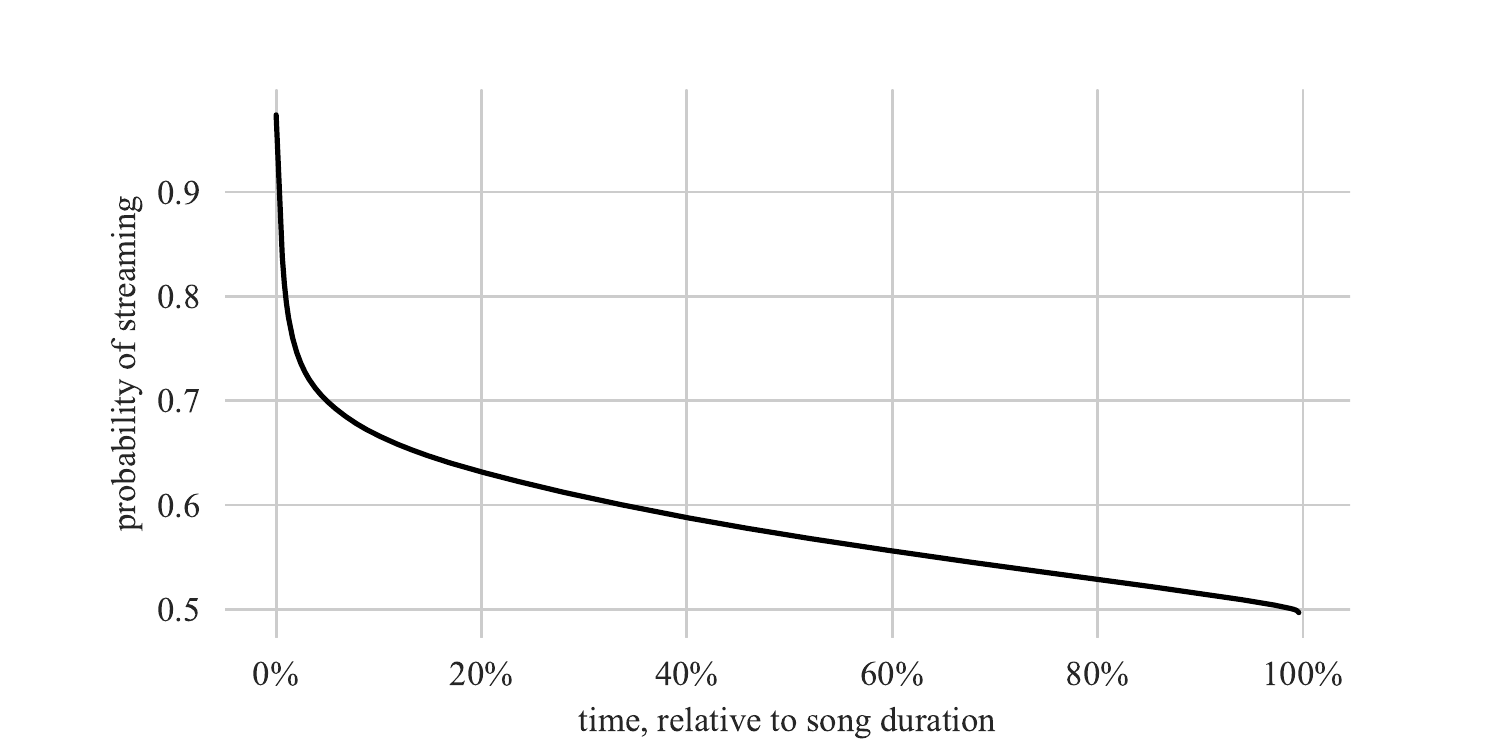}
    \caption{\label{fig:percentagestilllistening}
      Probability in time of a song being streaming.
    }
  \end{subfigure}

  \caption{\label{fig:skipspercentage} Collective skipping behavior of users.}

\end{figure}

Skipping is an overwhelmingly common behavior of users of streaming
services: a quarter of all streamed songs are skipped within the first
five seconds, and only roughly half of all songs are listened to in
their entirety~\cite{LamereSkips}. That analysis, which dates
back to 2014, was reproduced using song streams sampled in August 2018 from Spotify, and
its results were confirmed. As can be observed in Figure~\ref{fig:percentageprofile}
(in relative time) and~\ref{fig:skipprofile30s} (in absolute time),
most skips occur indeed at the very beginning of songs; there is also
a clear tendency to skip the ending of songs, which often contains
several seconds of silence or long fadeouts.
Figure~\ref{fig:percentagestilllistening} shows the (inverted)
cumulative distribution of skipping with respect to (normalized) time
(i.e., the probability in time that a song is still playing).

\subsection{Specificity and Consistency}

Figure~\ref{fig:specifitysnippet} shows sections of the Skip Profiles of
two songs, for which the data was collected on different days; this
particular example shows profiles that are unique to their respective
songs and consistent across collection dates.
This observation prompts the question of  whether
this is a general property that holds over a larger collection of
songs: can a song be identified from its Skip Profile? Subsequently,
the evolution of Skip Profiles is analyzed  considering data collected
over an extended period of consecutive days and
from different geographical regions.

\subsubsection*{Dataset}

The dataset used in this Section is composed of 100 popular songs,
released in April and May 2018. As of June 1st, 2018, 12 of those
songs were among the top 100 most popular songs (in terms of global
number of streams), and 40 of them among the top 1000. The songs were
selected by an expert musician, with the aim of maximizing variety
among genres and avoiding multiple songs from the same
artists. Over 3 billion skipping events have been collected from Spotify,
spanning a period of three months across all countries in which the
streaming service operates; most streaming activity (around 30\%)
originates from the United States, followed by Great Britain, Mexico,
Germany.

\subsubsection{Specificity of Skip Profiles}
\label{sec:specificitysp}

In order to study the specificity of the shape of Skip Profiles with respect to
songs, the dataset was processed by considering the first two minutes
of each profile (to account for the variability in length of songs),
and normalizing each resulting profile fragment independently (to
account for the different popularity of the songs); moreover, given
that most of the skipping activity occurs in the very first instants,
the initial 5 seconds of data are discarded as well.  Finally,
fragments are smoothed by median-filtering, to \mbox{de-noise}
profiles derived from smaller amounts of available streaming data.

\begin{figure}

  \centering

  \begin{subfigure}{0.9\textwidth}
    \centering
    \includegraphics[width=1\textwidth, trim={0.4cm 0.2cm 1.7cm 0},clip]{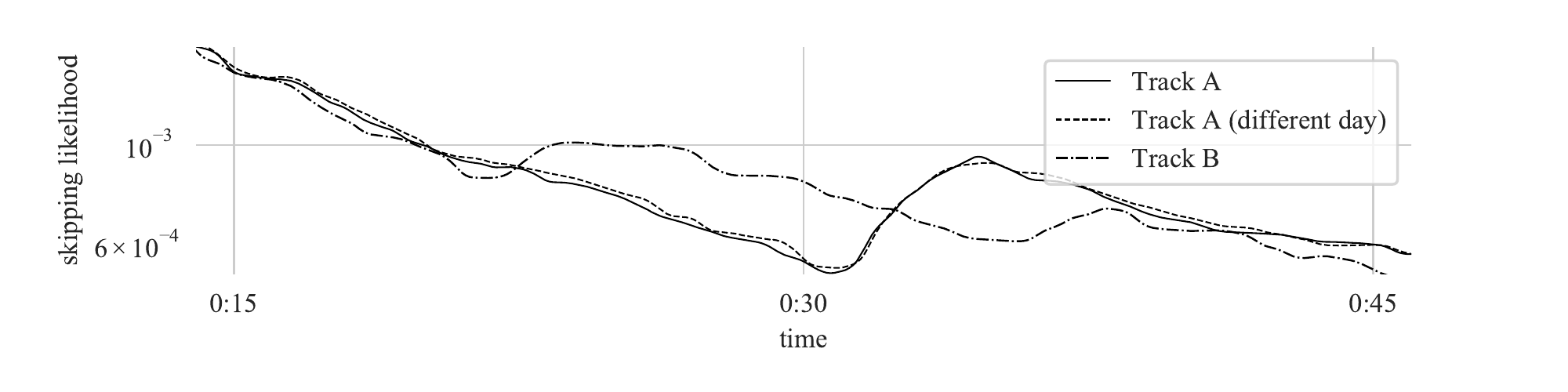}
  \caption{\label{fig:specifitysnippet} Sections of Skip Profiles collected on different days. }
  \end{subfigure}

  \begin{subfigure}{0.9\textwidth}
    \centering
    \includegraphics[width=1\textwidth, trim={0.9cm 0.2cm 1.7cm 0},clip]{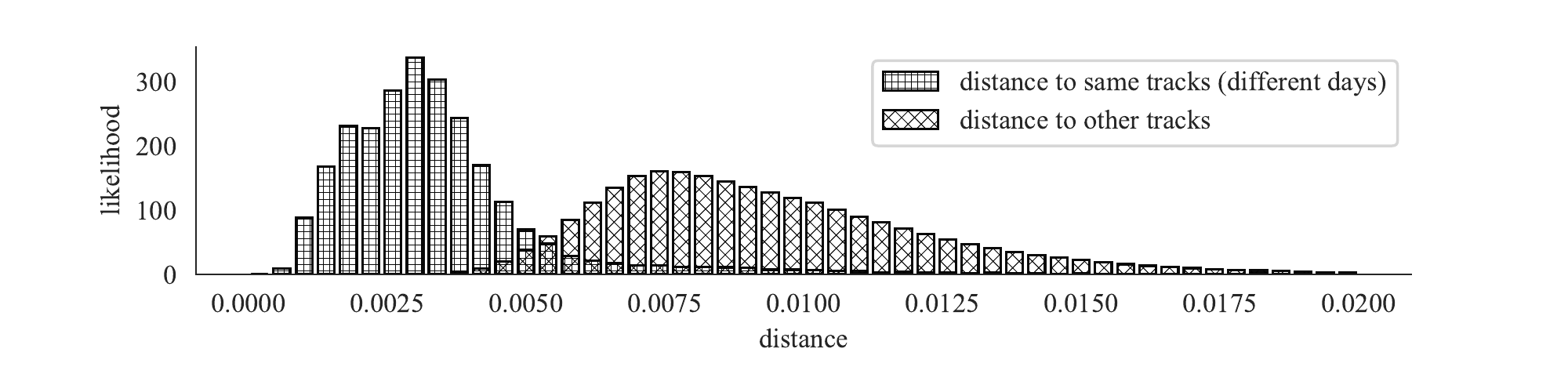}
  \caption{\label{fig:specifityhist} Distributions of same- and different-song distances.  }
  \end{subfigure}%

  \caption{Consistency of Skip Profiles in time. }

\end{figure}

Euclidean distance between the vector representation of Skip Profiles
fragments provides then a straightforward way to measure specificity:
profiles should ideally be closer to other profiles associated
to the same song (their streaming data being collected on different
days) than to profiles associated to different
songs. Figure~\ref{fig:specifityhist} shows that this is indeed the
case, by picturing the distributions of \mbox{same-song} and
\mbox{different-song} distances.

This specificity can be further quantified by framing the problem as
an Information Retrieval evaluation task~\cite{manningIRbook}.  Given
a query -- a skip profile for a random (track, date) pair --
the dataset is sorted by euclidean distance from the query.
Retrieved profiles are considered \emph{relevant} if associated to
the query track on different dates, \emph{non-relevant}
otherwise. Evaluation can then be carried out using common measures,
such as Mean Average Precision (MAP)\footnote{Precision is defined as
  the fraction of relevant documents among the retrieved ones. Average
  Precision is the average of the Precision values obtained
  considering the subsequences of the retrieved documents (sorted by
  relevance), up to each relevant document in the collection.  Mean
  Average Precision is the mean, over different queries, of the
  Average Precision value. The range for all these measures is
  $[0,1]$.}.

A random baseline for the experiment on this dataset (obtained by
returning a randomly shuffled list of results) yields \mbox{MAP =
0.014}; in contrast, the methodology discussed above yields \mbox{MAP = 0.886},
confirming the hypothesis of specificity of skip profiles.

\subsubsection{Evolution of Skip Profiles in Time}
\label{sec:evolutiontime}

Let us consider a signal consisting of the differences (in terms of
euclidean distance) between Skip Profiles collected on subsequent
days; several possibilities arise: a single S.P. can be selected as
reference (that corresponding to the first day being the obvious
choice), or the day-to-day difference between the subsequent days can
be examined.  In this section, the analysis is carried out on one
month of data collected following each song's release date; the
release dates on the Streaming service correspond to those of their
general availability.

\begin{figure}

  \begin{subfigure}{0.9\textwidth}
    \centering
    \includegraphics[width=0.95\textwidth]{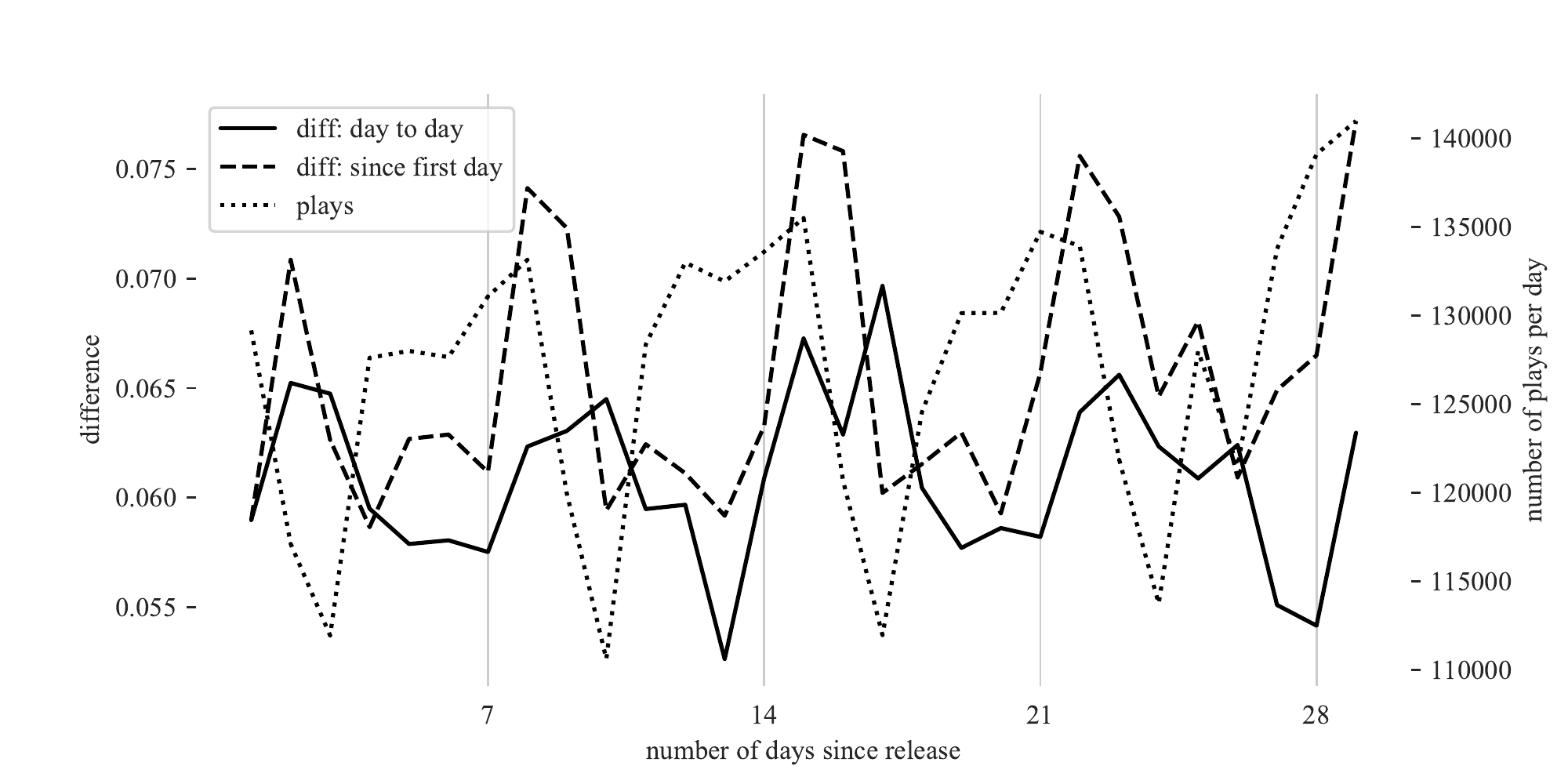}
    \caption{ \label{fig:evolution2} A song with a constant number of streams. }
  \end{subfigure}%

  \begin{subfigure}{0.9\textwidth}
    \centering
    \includegraphics[width=0.95\textwidth]{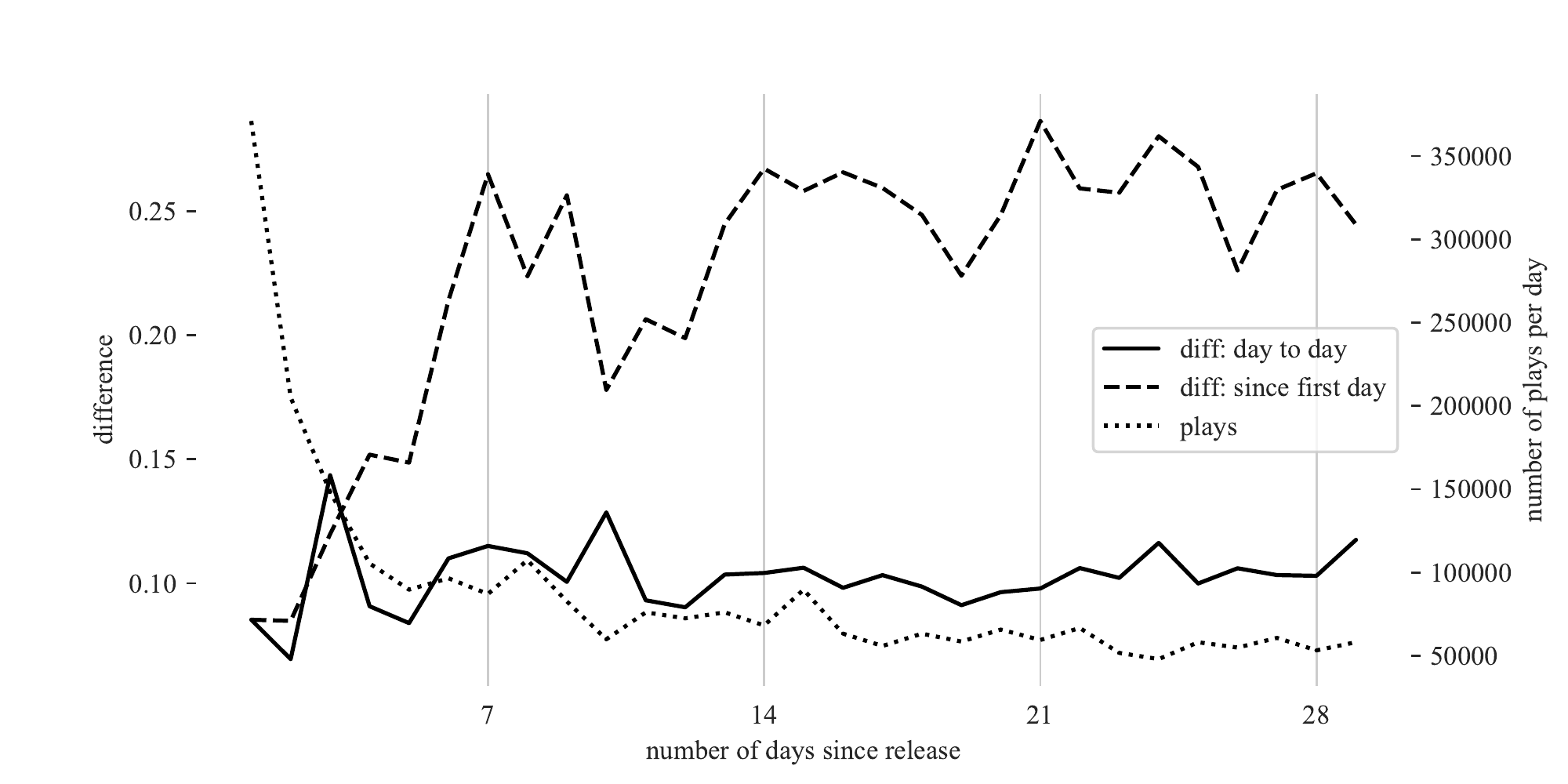}
    \caption{ \label{fig:evolution1} A song with a declining number of streams. }
  \end{subfigure}%

  \begin{subfigure}{0.99\textwidth}
    \centering
    \includegraphics[width=0.95\textwidth]{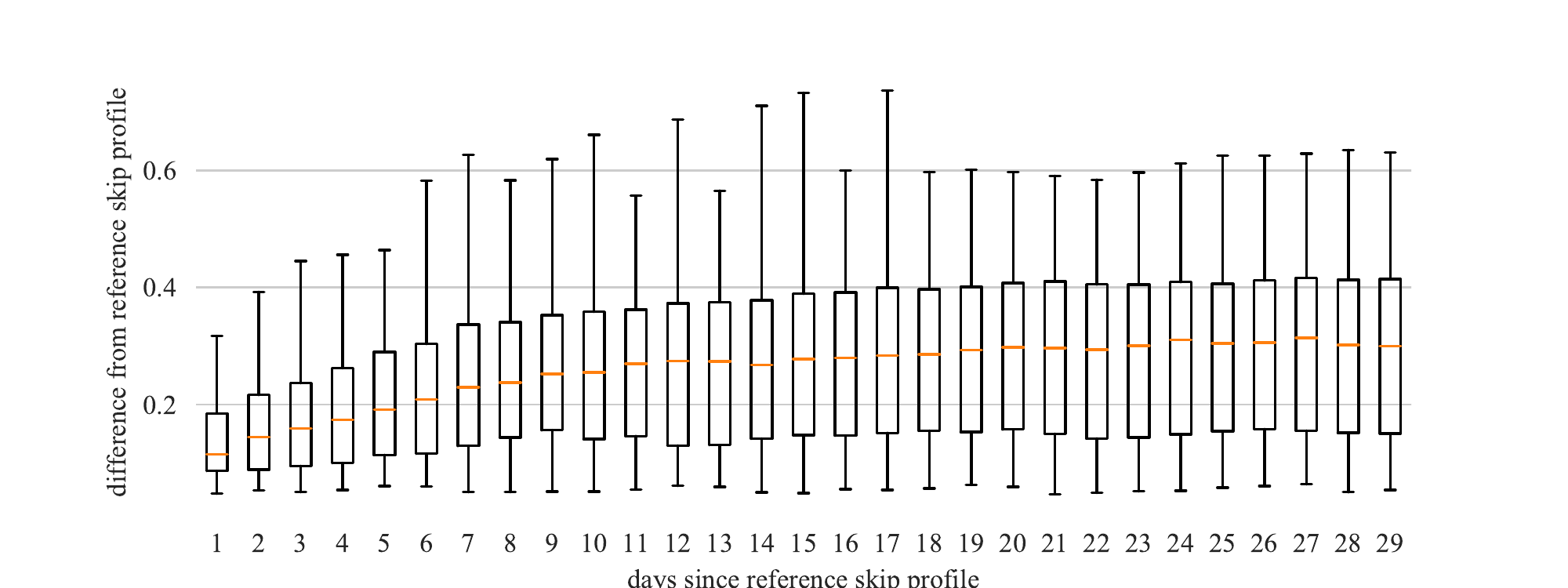}
    \caption{ \label{fig:evolutionboxplot} Change, with respect to the release date, over the dataset  }
  \end{subfigure}%

  \caption{ \label{fig:evolution} Difference between Skip Profiles collected on different days }

\end{figure}

Empirical analysis of the difference signal between subsequent days
showed no remarkable anomaly, i.e., there appears to be no point in
time in which user behaviour suddenly changes.  On the other hand, the
evolution of the difference signal with respect to the release date
shows more variability, and a determining factor seems to be
trajectory of the number of streams; Figure~\ref{fig:evolution}
exemplifies the two most common cases. A steady streaming behavior
(\ref{fig:evolution2}) is associated to a relatively constant
distribution of Skip Profiles over the different days (one can also
notice how the weekly listening patterns are reflected in the S.P.
evolution). On the other hand, the songs exhibiting significant
changes in S.P. over different days tend to be those for which the
amount of streams changes significantly: Figure~\ref{fig:evolution1}
shows a declining amount of streams per day, but the behavior for
rising amounts of streams is similar. Averaging over the dataset, one
can better sense the overall evolution:
Figure~\ref{fig:evolutionboxplot} shows, as one could reasonably
expect, that in the first two weeks the behavior undergoes the most
changes, after which Skip Profiles are fairly stable.

\subsubsection{Consistency of Profiles across Regions}

The geographical location of users presents an additional way of
subdividing user behaviour data. As was the case before, the object of
study is the consistency of Skip Profiles, across a different data
partitioning scheme.

Streaming data was collected for the same set of songs as in the
experiments presented above, and subdivided by country.  In order to
avoid being influenced by day of release, the data was collected in
August of 2018, to make sure that user behaviour for those songs is
stationary and that no effects caused by proximity to the
release date can be observed; Skip Profiles composed of less than
100.000 streams were discarded.

As anticipated, empirical observation of Skip Profiles collected
across different countries did not suggest any difference among
partitions. To validate this hypothesis, we repeated the experiments
of Section~\ref{sec:specificitysp}.  The resulting distribution of
\mbox{same- and} \mbox{different-song} distances is similar to the one
originating from the time-based partitioning; a MAP score of 0.939,
in contrast to a baseline of 0.017, was obtained, thus confirming the
hypothesis of consistency of profiles across data collection regions.

\section{User Behavior and Musical Structure}
\label{sec:structure}

\begin{figure}
  \begin{subfigure}{1\textwidth}
    \centering
    \includegraphics[width=1\textwidth]{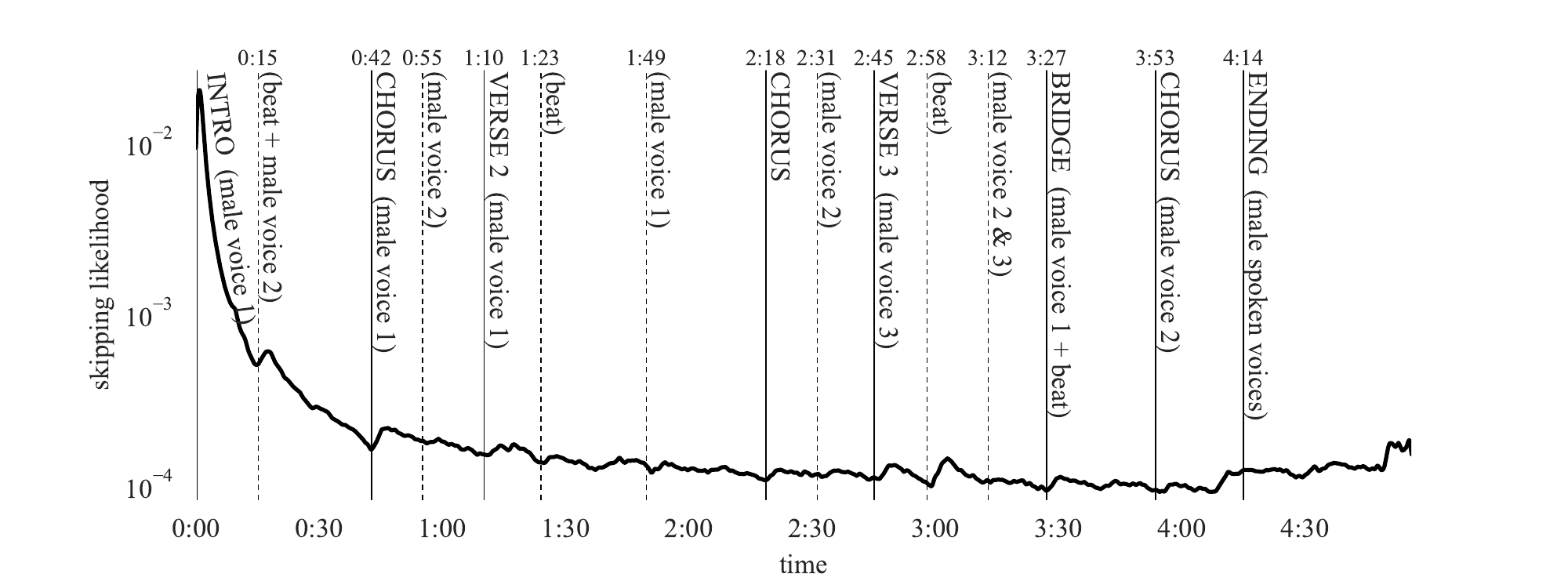}
    \caption{
      \label{fig:skipplusstructure}
      Skip Profile and annotated musical structural boundaries.
    }
  \end{subfigure}
  \begin{subfigure}{1\textwidth}
    \centering
    \includegraphics[width=1\textwidth]{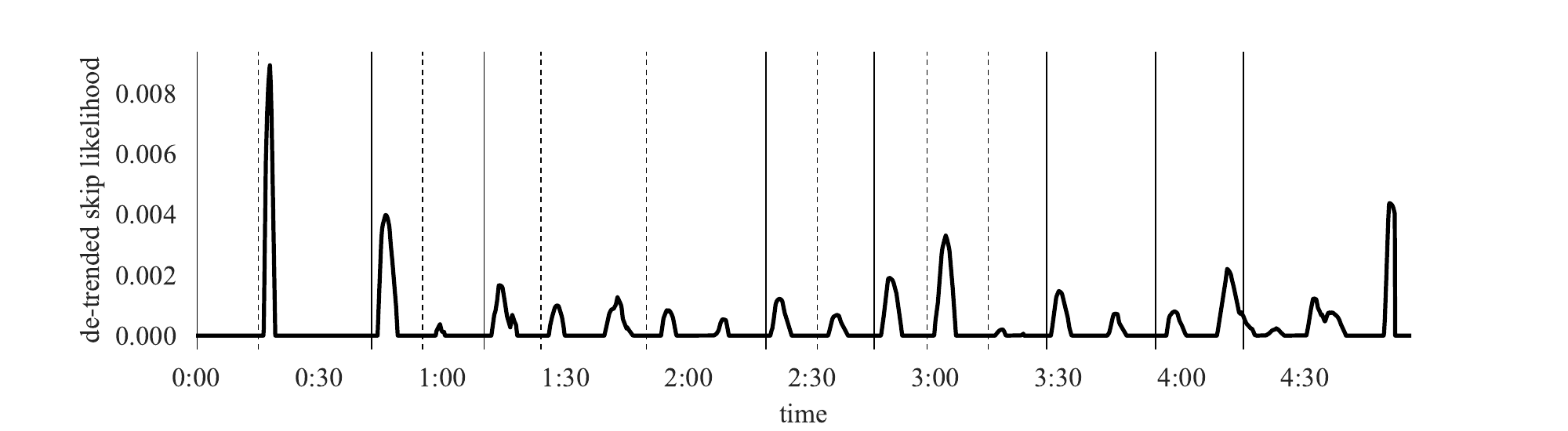}
    \caption{
      \label{fig:peaksplusstructure}
      De-trended Skip Profile and musical structural boundaries.
    }
  \end{subfigure}
  \begin{subfigure}{1\textwidth}
    \centering
    \includegraphics[width=1\textwidth]{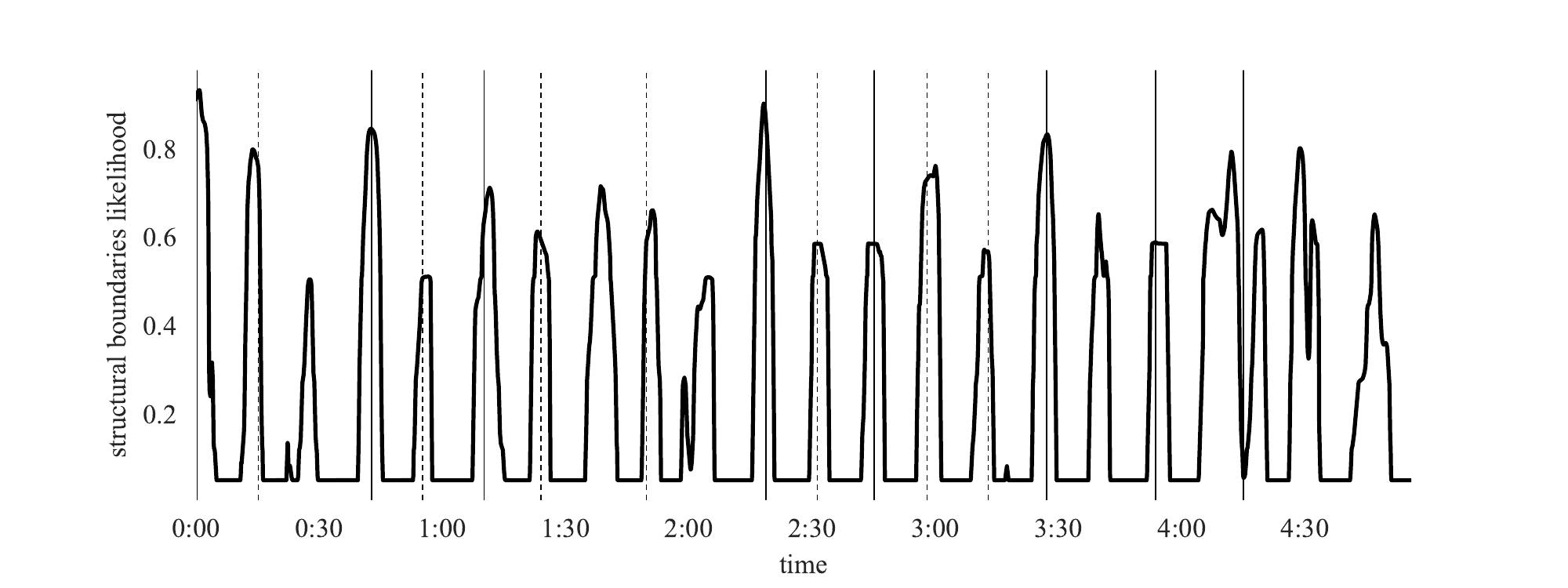}
    \caption{
      \label{fig:nnestplusstructure}
      Likelihood of structural boundaries, predicted from the Skip Profile.
    }
  \end{subfigure}
  \begin{subfigure}{1\textwidth}
    \centering
    \includegraphics[width=1\textwidth]{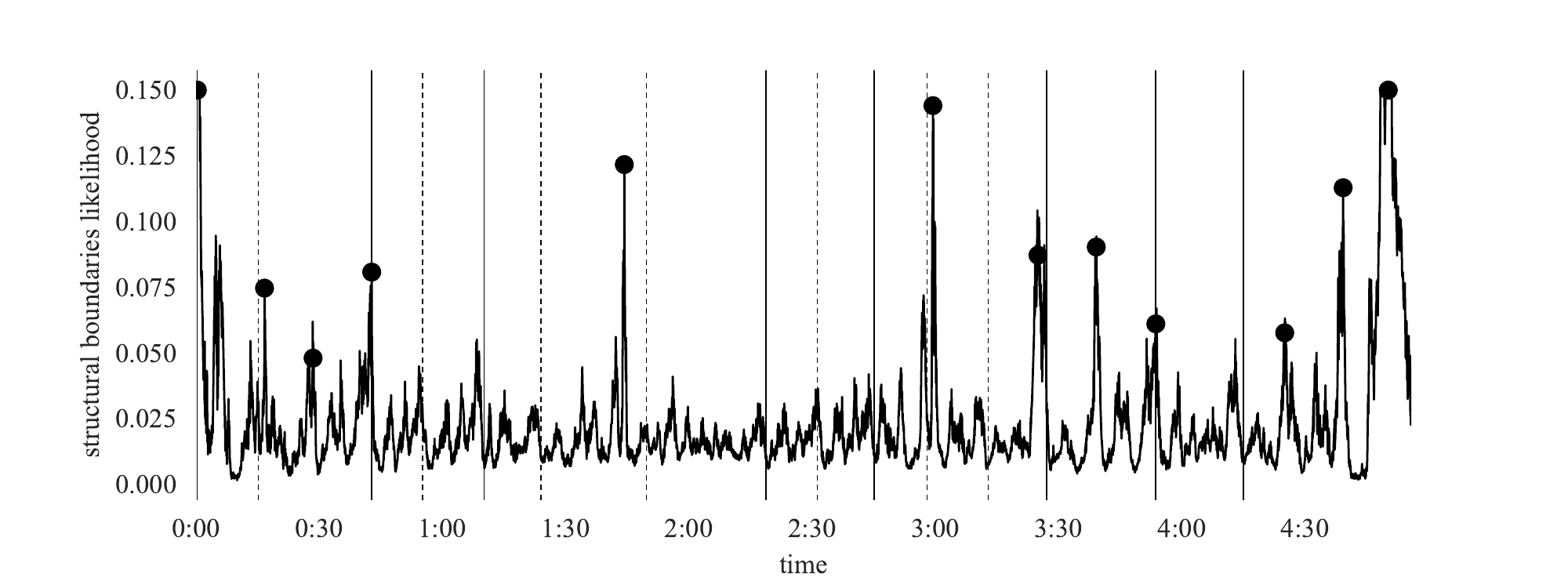}
    \caption{
      \label{fig:audioestplusstructure}
      Likelihood of structural boundaries, predicted from the acoustic content.
    }
  \end{subfigure}
  \caption{ Skipping behavior and musical structure.  }

\end{figure}

Individual Skip Profiles exhibit
deviations from their aggregated trend (\ref{fig:percentageprofile})
in a way that is closely related to musical structure.
Figure~\ref{fig:skipplusstructure} shows a song's Skip Profile,
overlayed with musical structural boundaries. It is visually evident
that section boundaries are commonly followed by surges in the
likelihood of skipping.

This Section investigates the correlation between user
skipping behavior and musical structure. To the best of the authors'
knowledge, such correlation has not been observed before.
It is first shown that the location of musically relevant boundaries
can be predicted directly from skip behavioral data. The accuracy of
the prediction is then evaluated against a collection of existing
algorithms, developed in the field of Music Information Retrieval,
that predict musical section boundaries from the content of recordings
(i.e., the acoustic waveforms).  Skipping behavior data is finally
exploited to automatically generate large quantities of training data
(otherwise very cumbersome and costly to create manually) for
\mbox{content-based} Machine Learning algorithms, and the accuracy of
algorithms trained using such data is evaluated against the customary
way of training, as well as multiple other baselines.

\subsection{Related Work}

\label{sec:audiorelatedwork}

\mbox{Content-based} musical structure segmentation has been a
central research subject in the field of Music Information Retrieval for
many years.  A comprehensive overview of the
topic~\cite{paulus2010audio} defines the main goal of a structural
segmentation algorithm: ``to divide an audio recording into temporal
segments corresponding to musical parts and to group these segments
into musically meaningful categories''. The input of such algorithms
is the audio content of a music recording (i.e., a digital encoding of
the acoustic waveform), the output is a partitioning of a song into
(usually) \mbox{non-overlapping} windows; some algorithms associate a
labeling to each section, which can be descriptive (such as ``bridge''
and ``chorus'') or just identifying of repetitive sections (e.g.,
``A'', ``B'' and so on). As discussed in the \mbox{above-mentioned}
paper, most algorithms can be categorized into three conceptual
approaches, based on \emph{repetition}, \emph{novelty} and
\emph{homogeneity}, that is the identification of, respectively,
recurring patterns, transitions between contrasting parts, and
contiguous sections that are consistent with respect to some musical
property; algorithms typically make use of features that attempt to
take into account musical characteristics such as melody, harmony,
rhythm, and timbre. A more recent overview of the State of the Art in
the field can be found in~\cite{nieto2015discovering}.

Research on the topic has long been hindered by the lack of sizeable
amounts of expert annotations -- the manual segmentations and labelings of
recordings done by \mbox{musically-competent} individuals -- because of its
\mbox{time-consuming} nature, coupled with the legal retrictions involved
in to sharing copyrighted recordings.  Initiatives such as the SALAMI
dataset~\cite{smith2011design} attempt to fill this need by providing
a relatively large (several hundreds) source of annotations for
recordings, many of which are in the public domain.

%% and beatles dataset TODO FIND CITATION

An alternative approach to deal with the aforementioned issues is the
release of the algorithms themselves as open source software:
\texttt{MSAF}~\cite{nieto2016systematic} is the leading effort in that regard
within the Music Information Retrieval community, and couples that goal with an
evaluation framework; the latter is in turn based on
\texttt{MIReval}~\cite{raffel2014mir_eval}, a reference implementation
of a large set of common \mbox{music-specific} IR evaluation metrics.
The remainder of this paper makes use of the following
algorithm implementations borrowed from \texttt{MSAF}:
 {\tt cnmf}~\cite{nieto2013convex},
 {\tt foote}~\cite{foote2000automatic},
 {\tt olda}~\cite{mcfee2014learning},
 {\tt scluster}~\cite{mcfee2014analyzing},
 {\tt sf}~\cite{serra2014unsupervised}.
Furthermore, another reference algorithm is given by the The Echo Nest
Analyzer, based on~\cite{jehan2005creating}, whose results can be
accessed through Spotify's public API.

Newer approaches, based on more recent Machine Learning techniques,
include \cite{ullrich2014boundary} and \cite{grill2015music}, that
achieve State of the Art accuracy in prediction by making use of
a Convolutional Neural Network architecture.

\subsection{Correlation of Skip Profiles and Musical Structure}
\label{sec:skips2struct}

In order to emphasize the regions of a Skip Profile that depart
significantly from its overall course, a de-trending algorithm is
applied, the results of which are depicted in
Figure~\ref{fig:peaksplusstructure}. The core of the procedure is a
deliberately simple heuristic (a combination of median and \mbox{low-pass}
filtering) aimed at isolating the general trend of the profile; this
trend is subsequently subtracted from the original signal and the
residual signal is rectified.

This graphical representation makes it even clearer
that surges in skips regularly follow section
boundaries, after a short delay.  Such delay (estimated to be
around 3.5s)  can be interpreted as the sum of two
components: the time it takes a user to realize they do not want to be
listening anymore to the current song -- triggered by the crossing of
a section boundary -- and the time spent interacting with the
reproduction device (tapping or clicking on the ``next'' button).

To prove the relation between user behavior and musical structure
we show that the latter (specifically, the location of section boundaries)
can be predicted from the former (a Skip Profile).
To this end, a compact Feed-Forward Neural Network (less than 50k
parameters) is trained on short segments (30s) of Skip Profiles; the
objective of the network is to predict whether the central location of
each particular segment in the original signal falls close enough
(within 1s) to a section boundary.  It is sufficient to annotate only
a few dozen songs to obtain satisfactory performance, and the results
of such procedure can be observed in
Figure~\ref{fig:nnestplusstructure}, clearly showing the strong
relation between music structure and user behavior (a more rigorous
quantitative evaluation is carried out in Section~\ref{sec:results}
below).

\subsection{Training of Acoustic Content-based Structure Prediction Models using Behavioral Data}
\label{sec:audio-nn-skips}
An existing algorithm from the Music Information Retrieval literature
is exploited to demonstrate the effectiveness of using Skip Profiles
for training \mbox{content-based} section prediction algorithms.

The algorithm, detailed in \cite{ullrich2014boundary}, is based on a straightforward,
well understood Convolutional Neural Network architecture. Originally
designed in \cite{schluter2014improved} for the task of \emph{onset
  detection} -- the task of detecting the instants at which musical
events, such as individual notes or chords, occur -- the
architecture of the model for structural prediction is unchanged, except for the longer input ranges
considered.
The model is essentially  a binary classifier that predicts the
likelihood of the presence of a section boundary in the center of an
audio excerpt.  The network comprises two convolutional layers, each
followed by a \mbox{max-pooling} layer, feeding one dense layer that
is finally projected into a scalar output.

The input of the network  is a segment of an
audio recording, for which a \mbox{\emph{Mel-Spectrogram}} is
extracted; the latter a common transformation in the music signal processing
literature \cite{muller2015fundamentals} consisting of a spectrogram (a \mbox{time-frequency}
representation obtained by concatenating the individual magnitude of
the Fourier transforms of short, overlapping excerpts of the signal)
whose frequency bands are then warped according to a perceptual (``Mel'') scale.

The  output is a scalar, whose value depends on the
distance of the closest section boundary from the center of the audio
input.  In \cite{ullrich2014boundary} a strategy known as \emph{target
  smearing} is employed, which accounts for the inaccuracy of ground
truth boundary annotations: during the training phase, only the excerpts centered on a frame that
is sufficiently close to a section boundary are  presented to the
network as positive examples, and their target value is the weight of
a Gaussian kernel, evaluated at the distance in time between the
center of the excerpt and the closest section boundary.

\mbox{Fine-grained} annotation of individual songs, in terms of the locations of
structural boundaries, is used for training this class of algorithms:
such annotation is carried out manually in a very laborious and
time consuming way.
In order to exploit of the correlation between Skip Profile and
structure, the model described in Section~\ref{sec:skips2struct} is
used to generate large quantities of training data. From the model
predictions, only regions of very high and very low boundary
likelihood are retained; empirical observation suggests that false
positive samples are generated  more frequently than false
negative ones, as can be seen in
Figure~\ref{fig:nnestplusstructure}.

The trained network can be used to create a prediction by repeatedly
applying it to adjacent, overlapping segments of a recording. The
output is a vector whose length is proportional to the length of the
input recording. In order to extract a discrete set of time instants
(the estimated structural boundaries) from it, a \mbox{peak-picking}
procedure is used: a point in the likelihood vector is considered a
peak if it is a local maximum and is far enough (a few seconds) from
other peaks; a threshold is set to half the value of the likelihood of
the \mbox{third-highest} peak.

An example prediction is pictured in
Figure~\ref{fig:audioestplusstructure}, along with the estimated
boundaries marked with circles.

\subsection{Evaluation}

The evaluation of segmentation algorithms is customarily formulated in
terms of (approximate) overlap between two sets of instants in time:
the actual timings of section boundaries (as annotated by an expert
curator) and those predicted by an algorithm.  Each prediction is
considered a \emph{hit} if it falls within a certain range from any
reference boundary timing, a \emph{miss} otherwise.

It is common in the literature to consider two such interval sizes, of
0.5 and 3s, and to use \mbox{F-score}\footnote{F-score is defined as the
  harmonic mean of Precision and Recall; the latter is defined as the
  fraction of relevant instances that have been retrieved over the
  total amount of relevant instances. }  as the preferred evaluation
measure. The evaluation of segmentation algorithms forms the subject
of \cite{nieto2014perceptual}, in which it is argued that an
appropriate weighting of the F-Score factors (as opposed to the
default unit weighting) is a measure that better corresponds to human
perception; the results presented below conform to such weighting
scheme.

\subsubsection*{Datasets}

The experiments were carried out making use of several datasets. No song belongs to multiple sets.

%% SALAMI STATS:
%% annotated songs: 1164
%% total annotations: 5784
%% annotated songs with audio: 376
%% total annotations with audio: 1770

\begin{itemize}

\item \texttt{SALAMI} \cite{smith2011design}: already mentioned in
  Section~\ref{sec:audiorelatedwork}, is comprised of the annotations
  for 1164 songs; however the recordings for only a subset of
  these songs are publicly available, yielding 376 useable
  recording/annotation pairs.  In case of multiple annotations per
  song, only the first one was considered.  These songs are not
  commercial recordings, hence no associated user behavior data is
  available.

\item \texttt{TOP100}: the dataset consists of the one hundred most
  popular songs (by number of streams, globally)
  on Spotify as of April 1, 2018.  The structure of each
  song was manually annotated by a single annotator (a professional
  musician).  The skip profiles for these songs, derived from roughly
  1 billion streams, were collected over the month of May 2018.

\item \texttt{SP20k}: a dataset of the Skip Profiles for roughly 20k
  songs; only songs with at least 100k streams over a period of 3
  months are retained, totalling about 81 billion streams.

\end{itemize}

In the \texttt{TOP100} dataset, two types of boundaries were
annotated: ``structural'', that only includes proper structural
boundaries (such as ``Intro'', ``Chorus'', ``Verse''), and
``extended'', that includes additional significant events happening in the
song (e.g., the entrance of a different singer within
a musical section).  In Figure~\ref{fig:skipplusstructure} the two
types of boundaries can be observed (solid line for structural
boundaries, dashed line for non-structural, extended boundaries).
Non-structural extended
boundaries often occur half-way through a musical section: a typical
case is a verse constituted by the repetition of a musical phrase, in
which the second occurrence is characterized by the presence of
additional (usually percussive) backing instruments.

\subsubsection*{Results}
\label{sec:results}

\definecolor{LightGray}{gray}{0.9}

\begin{table}
\centering
\begin{tabular}{lrrrr}
 \toprule
 Algorithm ~~~~~~~~~~~~~~~~~~~~~~~~~~~~~~~~~~~~~~~~ boundaries: & \multicolumn{2}{c}{Structural} & \multicolumn{2}{c}{Extended} \\
    ~~~~~~~~~~~~~~~~~~~~~~~~~~~~~~~~~~~~~~~~~~~~~~~~~~~~~~~ ``hit'' window size:                       & 0.5s & 3s & 0.5s & 3s \\
  \midrule
  \texttt{baseline}                                                     & 0.053 & 0.273 & 0.079 & 0.372 \\  \midrule
%   \texttt{cnmf~~~~~~~~~~~~~~~~~~~~} \cite{nieto2013convex}              & 0.056 & 0.277 & 0.095 & 0.392 \\
  \texttt{foote~~~~~~~~~~~~~~~~~~~} \cite{foote2000automatic}           & 0.085 & 0.415 & 0.127 & 0.511 \\
  \texttt{olda~~~~~~~~~~~~~~~~~~~~} \cite{mcfee2014learning}            & 0.158 & 0.504 & 0.217 & 0.609 \\
  \texttt{scluster~~~~~~~~~~~~~~~~} \cite{mcfee2014analyzing}           & 0.126 & 0.354 & 0.170 & 0.474 \\
  \texttt{sf~~~~~~~~~~~~~~~~~~~~~~} \cite{serra2014unsupervised}        & 0.106 & 0.425 & 0.165 & 0.502 \\
  \texttt{ten~~~~~~~~~~~~~~~~~~~~~} \cite{jehan2005creating}            & 0.152 & 0.536 & 0.190 & 0.607 \\
  \midrule
%   \rowcolor{LightGray}
  \texttt{skipprofile-nn~~~~~~~~~~} [Section~\ref{sec:skips2struct}]    & 0.245 & 0.630 & 0.278 & 0.636 \\
  \texttt{audio-nn-salami~~~~~~~~~} \cite{ullrich2014boundary}          & 0.226 & 0.464 & 0.287 & 0.522 \\ %  _trained_models-pskips_0.00-spvpos_1-spvneg_10-batch_64-rate_0.001-dropoutkeep_0.500-salamitwin_31-skipstwin_31-whitening_yes-schlueter_model.ckpt-50000
  \texttt{audio-nn-skip-profiles~~} [Section~\ref{sec:audio-nn-skips}]  & 0.259 & 0.560 & 0.307 & 0.638 \\ %  _trained_models-pskips_1.00-spvpos_1-spvneg_10-batch_64-rate_0.001-dropoutkeep_0.500-salamitwin_31-skipstwin_21-whitening_yes-schlueter_model.ckpt-50000
  \texttt{audio-nn-finetune~~~~~~~} [Section~\ref{sec:audio-nn-skips}]  & 0.311 & 0.575 & 0.373 & 0.658 \\ % _trained_models-pskips_0.00-spvpos_1-spvneg_10-batch_64-rate_0.000-dropoutkeep_1.000-salamitwin_31-skipstwin_31-whitening_yes_resumed-schlueter_model.ckpt-2500
 \bottomrule
\end{tabular}
\caption{\label{tab:mainresults} Hit Rate (weighted F-Score) for several algorithms, on the \texttt{TOP100} dataset.}
\end{table}

Table~\ref{tab:mainresults} reports the results of the evaluation,
in terms of Weighted F-Score for the \mbox{Hit-Rate} metric,
of several algorithms on the \texttt{TOP100} dataset.

The \texttt{baseline} entry refers to a trivial estimator, which
always predicts boundaries at fixed regular intervals; the particular
spacing (12s) has been determined through a \mbox{grid-search}
procedure, in order to obtain the highest possible \mbox{F-Score} for
the baseline. This particular baseline methodology mirrors that of
\cite{ullrich2014boundary}, in which the reported values are
however significantly higher (0.13 and 0.33, for window sizes of 0.5s
and 3s, respectively); the large difference is attributable to the different
evaluation dataset used, and to the weighting scheme applied to the
\mbox{F-Score} metric.

Next, the
% \texttt{cnmf},
\texttt{foote}, \texttt{olda}, \texttt{scluster}, and \texttt{sf}
entries refer to algorithms that have an open-source implementation
provided by
\texttt{MSAF}\footnote{\url{https://github.com/urinieto/msaf}}, and
\texttt{ten} refers to a commercial algorithm (\mbox{The Echo Nest},
now part of Spotify). In all of these instances, the default
parameters (if any) were utilized, and no training or optimization
procedure was performed.

The entry \texttt{skipprofile-nn} refers to the predictor described in
Section~\ref{sec:skips2struct} (due to the need to use the \texttt{TOP100}
dataset for this task, the reported values are computed through \mbox{5-fold}
\mbox{cross-validation}) and validates the fundamental thesis proposed
in this paper, namely, that musical structure and user behavior are
correlated.

The entry \texttt{audio-nn-salami} represents the model and training
procedure described in \cite{ullrich2014boundary}.  The original paper
presents several variants of the model, the largest of which was
adopted. Attempts were made to \mbox{re-implement} the described model as
closely as possible; because of the  smaller training set available to us
(this algorithm is trained using the publicly available, \mbox{376-songs}
subset of \texttt{SALAMI} as detailed above, as opposed to a dataset of
1220 annotated recordings available to the authors of the
original paper), and in order not to be hindered by possible misinterpretations
of unclear passages in the description,
several experiments were  performed
to find the best settings among variations of the original model and learning strategies.

The entry \texttt{audio-nn-skip-profiles} represents the same model of
\texttt{audio-nn-salami}, trained using data derived from user behavior
(the \texttt{SP20k} dataset) as described in
Section~\ref{sec:audio-nn-skips}.  The exact same parametrization and
learning strategy as before were used when performing this experiment, in order
to restrict the difference in outcome to the data source.  The
reported results show that it is indeed possible to achieve state of
the art performance using large quantities of
unannotated training data.

The final entry,
\texttt{audio-nn-finetune}, obtains the best performances by combining
both sources of training data.  The \texttt{audio-nn-skip-profiles}
model, discussed above, is taken as the starting point; subsequently,
it is \mbox{fine-tuned} by training the last layer using the
\texttt{SALAMI} dataset.  This allows the model to exploit both Skip
Profiles and manually assembled sources: the former, a source of
unreliable data available in large quantities, is used to learn a
robust feature representation, while the latter is used to
efficiently exploit the \mbox{high-precision} nature of a small,
\mbox{hand-curated} dataset.

\section{Discussion}

The original goal of the study was to leverage massive amounts of
\mbox{fine-grained} information collected by streaming services, in
order to better understand how songs are received by their audience. A
better understanding could in principle be used to inform the design
of novel compositional tools that take into account a model of the
listener grounded in actual music consumption data.

Through the investigation of \mbox{large-scale} user behavior, a
previously unknown correlation between skipping and musical structure
has emerged.  A joint analysis of the distribution of musical sections
and the overall trend of skip profiles, analyzed across a large
catalog (as opposed to with respect to individual songs), is the
object of future research inquiries. It is the authors' opinion that
such analysis has the potential to pave the way for the design of
tools that can leverage and anticipate user response in order to
provide useful guidance to creators.

An additional future research direction is the modeling of the
response of individual users to the songs to which they listen.  The
effect that musical features -- such as genre, mood, or
instrumentation, just to name a few -- have on the reception by users
is a well studied problem in the literature and in industry, and
closely tied to the field of Recommender Systems. The authors are however
not aware of existing work that attempts to jointly exploit \mbox{content-based} Music
Information Retrieval methods and user modeling
with \mbox{fine-grained} temporal user information to
influence recommendation, and believe
that this study can provide a starting point for such future research directions.

\subsubsection*{Acknowledgements}
The authors would like to acknowledge the contribution of Fernando Diaz for the
valuable insights during the initial phases of this work.

\bibliographystyle{alpha} \bibliography{main}

\end{document}